\documentclass[acmtocl,acmnow]{acmtrans2m}

\usepackage{epsfig}
\usepackage{latexsym}
\usepackage{xspace}
\usepackage{comment}
\usepackage{amsmath}
\usepackage{amssymb}

\newtheorem{theorem}{Theorem}[section]

\newtheorem{corollary}[theorem]{Corollary}

\newtheorem{lemma}[theorem]{Lemma}
\newdef{definition}[theorem]{Definition}
\newdef{remark}[theorem]{Remark}
\newdef{example}[theorem]{Example}

\newcommand{\fpbox}[3]{\framebox[#1]{\parbox{#2}{#3}}}
\newcommand{\boxtext}[1]{\fpbox{\textwidth}{0.9\textwidth}{#1}}

\newcommand{\If}{\lq\lq$\Leftarrow$\rq\rq\ \ }           
\newcommand{\OnlyIf}{\lq\lq$\Rightarrow$\rq\rq\ \ }      

\newcommand{\C}{\mathcal{C}} \newcommand{\D}{\mathcal{D}}

\newcommand{\I}{\mathcal{I}} 
 
\newcommand{\M}{\mathcal{M}}

\renewcommand{\S}{\mathcal{S}} \newcommand{\T}{\mathcal{T}}

\newcommand{\Ra}{\Rightarrow}
\newcommand{\la}{\leftarrow}

\newcommand{\lora}{\longrightarrow}

\newcommand{\cld}{,\ldots,}		     
\newcommand{\per}{\mbox{\bf .}}   
\newcommand{\ol}[1]{\overline{#1}}                   
\newcommand{\dd}[2]{#1_1,\ldots,#1_{#2}}	
\newcommand{\card}[1]{\##1}
\newcommand{\set}[1]{\{#1\}}
\newcommand{\setone}[2][1]{\set{#1\cld #2}}
\newcommand{\incl}{\subseteq}

\newcommand{\dom}[1][\I]{\Delta^{#1}}            
\newcommand{\Int}[2][\I]{#2^{#1}}                
\newcommand{\INT}[2][\I]{(#2)^{#1}}              
\newcommand{\inter}[1][\I]{(\dom[#1],\Int[#1]{\cdot})}   

\newcommand{\ISA}{\sqsubseteq}

\newcommand{\AND}{\sqcap}
\newcommand{\OR}{\sqcup}
\newcommand{\NOT}{\neg}

\newcommand{\ALL}[2]{\forall #1 \per #2}
\newcommand{\SOME}[2]{\exists #1 \per #2}
\newcommand{\SOMER}[2]{\exists[#1]#2}
\newcommand{\ATMOSTR}[3]{(\leq #1\, [#2]#3)}
\newcommand{\COMP}[2]{(#1\mathop{:}#2)}
\newcommand{\PROJ}[3]{#1|_{#2,#3}}

\newcommand{\true}{\mathsf{T}}
\newcommand{\false}{\mathsf{F}}
\newcommand{\limp}{\mathbin{\Ra}}
\newcommand{\DIAM}[2]{\langle #1 \rangle #2}
\newcommand{\BOX}[2]{[#1]#2}
\newcommand{\QBOX}[3]{[#2]_{\leq #1}#3}

\newcommand{\rR}{\mathbf{R}}
\newcommand{\pP}{\mathbf{P}}
\newcommand{\uu}{U}

\newcommand{\CIQ}{\ensuremath{\mathcal{CIQ}}\xspace}

\newcommand{\DLRreg}{\ensuremath{\mathcal{DLR}_\mathit{\MakeLowercase{reg}}}\xspace}

\newcommand{\CPDLg}{\textsc{cpdl}\ensuremath{_{\MakeLowercase{g}}}\xspace}

\newcommand{\CL}{\mathit{CL}}
\newcommand{\Pre}{\mathit{Pre}}

\newcommand{\conj}{\mathit{conj}}

\newcommand{\vett}[1]{\vec{\mathbf{#1}}}
\newcommand{\create}{\mathit{create}}

\newcommand{\conjl}[1][j]{\conj_{#1}(\vett{a},\vett{b}_{#1},\vett{c}_{#1})}

\newcommand{\conjr}[1][j]{\exists\vett{z}_{#1}\per
                          \conj'_{#1}(\vett{a},\vett{z}_{#1},\vett{c}_{#1})}
\newcommand{\conjpart}{\exists\vett{w}_{\pi}\per
                       \conj'_j(\vett{a},\vett{w}_{\pi},\vett{c})}
\newcommand{\deltat}[1][]{\delta^{\pi}_{#1}}

\newcommand{\cert}[3]{\mathit{cert}(#1,#2,#3)}

\newcommand{\nmax}{n_{\mathit{max}}}

\newcommand{\Phisqq}{\Phi_{S\not\models q\subseteq q'}}
\newcommand{\Phiaux}{\Phi_{\mathit{aux}}}

\newcommand{\Nat}{{\mathbb N}}

\newcommand{\Tile}{\mathit{Tile}}
\newcommand{\Right}{\mathit{Right}}
\newcommand{\Up}{\mathit{Up}}

\newcommand{\CONTROLS}{\texttt{CONTROLS}}
\newcommand{\SOLD}{\texttt{SOLD}}
\newcommand{\Dept}{\texttt{Dept}}
\newcommand{\MainDept}{\texttt{MainDept}}
\newcommand{\Company}{\texttt{Company}}
\newcommand{\Money}{\texttt{Money}}

\title{Conjunctive Query Containment and Answering under Description Logics
 Constraints$^*$}

\author{DIEGO CALVANESE\\
 Faculty of Computer Science\\
 Free University of Bozen-Bolzano\\
 Piazza Domenicani 3, 39100 Bolzano, Italy\\
 \texttt{calvanese@inf.unibz.it}\\
 ~\\
 GIUSEPPE DE GIACOMO, and MAURIZIO LENZERINI\\
 Dipartimento di Informatica e Sistemistica\\
 Universit\`a di Roma ``La Sapienza''\\
 Via Salaria 113, 00198 Roma, Italy\\
 \{\texttt{degiacomo},\texttt{lenzerini}\}\texttt{@dis.uniroma1.it}
 }

\begin{abstract}
Query containment and query answering are two important computational
tasks in databases. While query answering amounts to compute the
result of a query over a database, query containment is the problem of
checking whether for every database, the result of one query is a
subset of the result of another query.

In this paper, we deal with unions of conjunctive queries, and we
address query containment and query answering under Description Logic
constraints. Every such constraint is essentially an inclusion
dependencies between concepts and relations, and their expressive
power is due to the possibility of using complex expressions, e.g.,
intersection and difference of relations, special forms of
quantification, regular expressions over binary relations, in the
specification of the dependencies. These types of constraints capture
a great variety of data models, including the relational, the
entity-relationship, and the object-oriented model, all extended with
various forms of constraints, and also the basic features of the
ontology languages used in the context of the Semantic Web.

We present the following results on both query containment and query
answering.  We provide a method for query containment under
Description Logic constraints, thus showing that the problem is
decidable, and analyze its computational complexity. We prove that
query containment is undecidable in the case where we allow
inequalities in the right-hand side query, even for very simple
constraints and queries. We show that query answering under
Description Logic constraints can be reduced to query containment, and
illustrate how such a reduction provides upper bound results with
respect to both combined and data complexity.
\end{abstract}

\category{...}{...}{...}
\terms{Algorithms, Theory}
\keywords{...}

\begin{document}

\begin{bottomstuff}
  $^*$ This paper is an extended and revised version of a paper published in
  the Proceedings of the 17th ACM SIGACT SIGMOD SIGART Sym.\ on Principles of
  Database Systems (PODS'98).\\
\end{bottomstuff}

\maketitle


\section{Introduction}
\label{sec-introduction}

Query containment and query answering are two important computational
tasks in databases. While query answering amounts to compute the
result of a query over a database, query containment is the problem of
checking whether for every database, the result of one query is a
subset of the result of another query%
\footnote{We refer to the set semantics of query containment. Bag
  semantics is studied, for example, in \cite{IoRa95}.}. %
Many papers point out that checking containment is a relevant task in
several contexts, including information integration \cite{Ullm97},
query optimization \cite{AbHV95,AhSU79}, (materialized) view
maintenance \cite{GuMu95}, data warehousing \cite{Wido95b}, constraint
checking \cite{GSUW94}, and semantic caching \cite{APTP03}.

In this paper, we deal with query containment and query answering
under integrity constraints, or simply constraints. 

The former is the problem of checking whether containment between two
queries holds for every database satisfying a given set of
constraints.  This problem arises in those situation where one wants
to check query containment relatively to a database schema specified
with a rich data definition language.  For example, in the case of
information integration, queries are often to be compared relatively
to (inter-schema) constraints, which are used to declaratively specify
the ``glue'' between two source schemas, and between one source schema
and the global schema
\cite{CDLNR98,Hull97,Ullm97,CaLe93,LeSK95,Lenz02,Hale01}.

The complexity of query containment in the absence of constraints has
been studied in various settings.  In \cite{ChMe77}, NP-completeness
has been established for conjunctive queries, and in \cite{ChRa97} a
multi-parameter analysis has been performed for the same case, showing
that the intractability is due to certain types of cycles in the
queries.  In \cite{Klug88,Meyd98}, $\Pi^p_2$-completeness of
containment of conjunctive queries with inequalities was proved, and
in \cite{SaYa80} the case of queries with the union and difference
operators was studied.  For various classes of Datalog queries with
inequalities, decidability and undecidability results were presented
in \cite{ChVa92,Meyd98,Bona04,CaDV03}, respectively.

Query containment under constraints has also been the subject of
several investigations.  For example, decidability of conjunctive
query containment was investigated in \cite{AhSU79b} under functional
and multi-valued dependencies, in \cite{JoKl84} under functional and
inclusion dependencies, in \cite{Chan92,LeRo96d,LeSu97} under
constraints representing \emph{is-a} hierarchies and complex objects,
and in \cite{DoSu96} in the case of constraints represented as Datalog
programs. Undecidability is proved in \cite{CaRo03} for recursive
queries under inclusion dependencies. Several results on containment
of XML queries under constraints expressed as DTDs are reported in
\cite{NeSc03,Wood03}.

Query answering under constraints is the problem of computing the
answers to a query over an incomplete database relatively to a set of
constraints \cite{Meyd98}. Since an incomplete database is partially
specified, this task amounts to compute the tuples that satisfy the
query in every database that conforms to the partial specification,
and satisfies the constraints. It is well known in the database
literature that there is a tight connection between the problems of
conjunctive query containment and conjunctive query answering
\cite{ChMe77}. Since this relationship holds also in the presence of
constraints, most of the results reported above apply to query
answering as well. In this paper, we concentrate mainly on query
containment, and address query answering only in
Section~\ref{sec-answering}.

\smallskip

In this paper\footnote{This paper is an improved and extended version
  of part of \cite{CaDL98}.}, we address query containment and
answering in a setting where:
\begin{longitem}
\item The schema is constituted by concepts (unary relations) and
  relations as basic elements, and by a set of constraints expressed
  in a variant of Description Logics \cite{BCMNP03}.  Every constraint
  is an inclusion of the form $\alpha_1\incl\alpha_2$, where
  $\alpha_1$ and $\alpha_2$ are complex expressions built by using
  intersection and difference of relations, special forms of
  quantification, regular expressions over binary relation, and number
  restrictions (i.e.~cardinality constraints imposing limitations on
  the number of tuples in a certain relation in which an object may
  appear).  The constraints express essentially inclusion dependencies
  between concepts and relations, and their expressive power is due to
  the possibility of using complex expressions in the specification of
  the dependencies.  It can be shown that our formalism is able to
  capture a great variety of data models, including the relational,
  the entity-relationship, and the object-oriented model, all extended
  with various forms of constraints. The relevance of the constraints
  dealt with in this paper is also testified by the large interest
  that the Semantic Web community expresses towards Description
  Logics.  Indeed, several papers point out that ontologies play a key
  role in developing Semantic Web tools \cite{Grub93}, and Description
  Logics are regarded as the main formalisms for the specification of
  ontologies in this context \cite{PaHH04}. Despite this interest, the
  results presented in this paper can be considered one of the first
  formal analysis on querying ontologies.

\item Queries are formed as disjunctions of conjunctive queries whose
  atoms are concepts and relations, and therefore can express
  non-recursive Datalog programs.

\item An incomplete database is specified as a set of facts asserting
  that a specific object is an instance of a concept, or that a
  specific tuple of objects is an instance of a relation. As we said
  before, an incomplete database $\D$ is intended to provide a partial
  specification of a database, in the sense that a database conforming
  to $\D$ contains all facts explicitely asserted in $\D$, and may
  contain additional intances of concepts and relations.
\end{longitem}

We observe that, given the form of constraints and queries allowed in
our approach, none of the previous results can be applied to get
decidability/undecidability of query containment and query answering
in our setting.

We present the following results on both query containment and query answering:
\begin{enumerate}
\item We provide a method for query containment under Description
  Logic constraints, thus showing that the problem is decidable, and
  analyze its computational complexity. This result is obtained by
  adopting a novel technique for addressing the problem, based on
  translating the schema and the queries into a particular
  Propositional Dynamic Logic (PDL) formula, and then checking the
  unsatisfiability of the formula.  The technique is justified by the
  fact that reasoning about the schema itself (without the queries) is
  optimally done within the framework of PDL \cite{DeLe96}.
\item We prove that query containment is undecidable in the case where we allow
  inequalities in the right-hand side query, even for very simple constraints
  and queries.
\item We show that query answering under Description
  Logic constraints can be reduced to query containment, and
  illustrate how such a reduction provides upper bound results with
  respect to both combined and data complexity.
\end{enumerate}

The paper is organized as follows.  In Section~\ref{sec-formalism}, we present
the formalism used to express both the constraints in the schema, and the
queries.  In Section~\ref{sec-containment}, we deal with query containment.  In
particular, in Subsection~\ref{sec-cpdlg} we describe the logic \CPDLg, which
will be used for deciding query containment, in Subsection~\ref{sec-reduction}
we describe the reduction of query containment to unsatisfiability in \CPDLg,
in Subsection~\ref{sec-correctness} we prove its correctness, and in
Section~\ref{sec-complexity} we analyze the complexity bounds for checking
containment of queries.  In Section~\ref{sec-undecidability}, we show
undecidability of query containment in the presence of inequalities.  In
Section~\ref{sec-answering}, we deal with query answering, and in
Section~\ref{sec-conclusions} we conclude the paper.



\section{Schemas and Queries in \DLRreg}
\label{sec-formalism}

To specify database schemas and queries, we use the logical language \DLRreg,
inspired by \cite{CaLe93,CaDL95}, belonging to the family of (expressive)
Description Logics \cite{CDLN01,BCMNP03}.  The language is based on the
relational model, in the sense that a schema $\S$ describes the properties of a
set of relations, while a query for $\S$ denotes a relation that is supposed to
be computed from any database conforming to $\S$. A schema is specified in
terms of a set of assertions on relations, which express the constraints that
must be satisfied by every conforming database.

\subsection{Schemas}

The basic elements of \DLRreg are \emph{concepts} (unary relations),
\emph{$n$-ary relations}, and \emph{regular expressions} built over projections
of relations on two of their components.%
\footnote{We could include in the logic also domains, i.e.~sets of values such
 as integer, string, etc..  However, for the sake of simplicity, we
 do not consider this aspect in this work.}. %

We assume to deal with a finite set of atomic concepts and relations, denoted
by $A$ and $\pP$ respectively.  We use $C$ to denote arbitrary concepts, $\rR$
to denote arbitrary relations (of given arity between~2 and $n_{max}$), and $E$
to denote regular expressions, respectively built according to the following
syntax
\begin{align*}
  C   &::= \top_1 ~\mid~ A ~\mid~ \NOT C ~\mid~ C_1\AND C_2 ~\mid~
           \SOME{E}{C} ~\mid~
           \SOMER{\$i}{\rR} ~\mid~ \ATMOSTR{k}{\$i}{\rR}\\
  \rR &::= \top_n ~\mid~ \pP ~\mid~ \COMP{\$i/n}{C} ~\mid~ \NOT \rR ~\mid~
           \rR_1\AND\rR_2\\
  E   &::= \varepsilon ~\mid~ \PROJ{\rR}{\$i}{\$j} ~\mid~ E_1\circ E_2 ~\mid~
           E_1\OR E_2 ~\mid~ E^*
\end{align*}
where $i$ and $j$ denote components of relations, i.e., integers between~1 and
$n_{max}$, $n$ denotes the arity of a relation, i.e., an integer between~2 and
$n_{max}$, and $k$ denotes a nonnegative integer.

Expressions of the form $\ATMOSTR{k}{\$i}{\rR}$ are called \emph{number
 restrictions}.  In what follows, we abbreviate $\NOT\SOME{E}{\NOT C}$ with
$\ALL{E}{C}$, and $\COMP{\$i/n}{C}$ with $\COMP{\$i}{C}$ when $n$ is clear from
the context.  Also, we consider only concepts and relations that are
\emph{well-typed}, which means that
\begin{itemize}
\item only relations of the same arity $n$ are combined to form expressions of
  type $\rR_1\AND\rR_2$ (which inherit the arity~$n$), and
\item $i\leq n$ whenever $i$ denotes a component of a relation of arity $n$.
\end{itemize}

A \DLRreg \emph{schema} is constituted by a finite set of \emph{assertions}, of
the form
\begin{align*}
  C_1   &\ISA C_2\\
  \rR_1 &\ISA \rR_2
\end{align*}
where $\rR_1$ and $\rR_2$ are of the same arity.  Note that our notion of
schema corresponds to that of TBox in Description Logics \cite{BCMNP03}.

The semantics of \DLRreg is specified through the notion of interpretation.  An
\emph{interpretation} $\I=\inter$ of a \DLRreg schema $\S$ and a set $\C$ (of
constants to be used in queries) is constituted by an \emph{interpretation
 domain} $\dom$ and an \emph{interpretation function} $\Int{\cdot}$ that
assigns
\begin{itemize}
\item to each constant $c$ in $\C$ an element $\Int{c}$ of $\dom$ under the
  unique name assumption,
\item to each concept $C$ a subset $\Int{C}$ of $\dom$,
\item to each relation $\rR$ of arity $n$ a subset $\Int{\rR}$ of $(\dom)^n$,
\item to each regular expression $E$ a subset $\Int{E}$ of $\dom\times\dom$
\end{itemize}
such that the conditions in Figure~\ref{fig-semantics} are satisfied.  We
observe that $\top_1$ denotes the interpretation domain, while $\top_n$, for
$n>1$, does \emph{not} denote the $n$-Cartesian product of the domain, but only
a subset of it, that covers all relations of arity~$n$.  It follows from this
property that the ``$\NOT$'' constructor on relations is used to express
difference of relations, rather than complement.

\begin{figure}[t]
  \boxtext{
   \[
     \begin{array}{rcl}
       \Int{\top_1} &=& \dom\\
       \Int{A} &\incl& \dom\\
       \INT{\NOT C} &=& \dom\setminus\Int{C}\\
       \INT{C_1\AND C_2} &=& \Int{C_1}\cap\Int{C_2}\\
       \INT{\SOME{E}{C}} &=&
         \set{d\in\dom \mid \exists d'\in\Int{C} \per (d,d')\in\Int{E}}\\
       \INT{\SOMER{\$i}{\rR}} &=&
         \set{d\in\dom \mid \exists(\dd{d}{n})\in\Int\rR \per d_i=d}\\
       \Int{\ATMOSTR{k}{\$i}{\rR}} &=&
         \set{d\in\dom \mid
            \card{\set{(\dd{d}{n})\in\Int\rR_1 \mid d_i=d}} \leq k}\\[3mm]
       \Int{\top_n} &\incl& (\dom)^n\\
       \Int{\pP} &\incl& \Int{\top_n}\\
       \Int{\COMP{\$i/n}{C}} &=&
         \set{(\dd{d}{n})\in\Int{\top_n}\mid d_i\in\Int{C}}\\
       \INT{\NOT\rR} &=& \Int{\top_n}\setminus\Int{\rR}\\
       \INT{\rR_1\AND\rR_2} &=& \Int{\rR_1}\cap\Int{\rR_2}\\[3mm]
       \Int{\varepsilon} &=& \set{(x,x) \mid x\in\dom}\\
       \INT{\PROJ{\rR}{\$i}{\$j}} &=&
         \set{(x_i,x_j) \mid (\dd{x}{n})\in\Int{\rR}}\\
       \INT{E_1\circ E_2} &=& \Int{E_1}\circ\Int{E_2}\\
       \INT{E_1\OR E_2} &=& \Int{E_1}\cup\Int{E_2}\\
       \INT{E^*} &=& (\Int{E})^*
     \end{array}
   \]
   }
  \caption{Semantic rules for \DLRreg ($\pP$, $\rR$, $\rR_1$, and $\rR_2$
   have arity $n$)}
  \label{fig-semantics}
\end{figure}

An interpretation $\I$ \emph{satisfies} an assertion $C_1\ISA C_2$ (resp.,
$\rR_1\ISA\rR_2$) if $\Int{C_1}\incl\Int{C_2}$ (resp.,
$\Int{\rR_1}\incl\Int{\rR_2}$).  An interpretation that satisfies all
assertions in a schema $\S$ is called a \emph{model} of $\S$.  It is easy to
see that a model of a schema $\S$ actually corresponds to a database conforming
to $S$, i.e., a database satisfying all the constraints represented by $\S$.  A
schema is \emph{satisfiable} if it admits a model.  A schema $\S$
\emph{logically implies} an inclusion assertion $C_1\ISA C_2$
(resp.~$\rR_1\ISA\rR_2$) if for every model $\I$ of $\S$ we have that
$\Int{C_1}\incl\Int{C_2}$ (resp.~$\Int{\rR_1}\incl\Int{\rR_2}$).

\smallskip

It can be shown that \DLRreg is able to capture a great variety of data models
with many forms of constraints.  For example, we obtain the entity-relationship
model (including \emph{is-a} relations on both entities and relations) in a
straightforward way \cite{CaDL95}, and an object-oriented data model (extended
with several types of constraints), by restricting the use of existential and
universal quantifications in concept expressions, by restricting the attention
to binary relations, and by eliminating negation, disjunction and regular
expressions. Compared with the relational model, the following observations
point out the kinds of constraints that can be expressed using \DLRreg.
\begin{itemize}
\item Assertions directly express a special case of typed inclusion
  dependencies, namely the one where no projection of relations is used.
\item Unary inclusion dependencies are easily expressible by means of the
  $\SOMER{\$2}{\pP}$ construct. For example,
  $\SOMER{\$2}{\pP_1}\ISA\SOMER{\$3}{\pP_2}$ is a unary inclusion dependency
  between attribute~2 of $\pP_1$ and attribute~3 of $\pP_2$.
\item Existence and exclusion dependencies are expressible by means of
  $\exists$ and $\neg$, respectively, whereas a limited form of functional
  dependencies can be expressed by means of $\ATMOSTR{1}{\$i}{\rR}$. For
  example, $\top_1 \ISA \ATMOSTR{1}{\$i}{\pP}$ specifies that attribute $i$
  functionally determines all other attributes of $\pP$.
\item The possibility of constructing complex expressions provides a special
  form of view definition. Indeed, the two assertions $\pP\ISA\rR$,
  $\rR\ISA\pP$ (where $\rR$ is a complex expression) is a view definition for
  $\pP$. Notably, views can be freely used in assertions (even with cyclic
  references), and, therefore, all the above discussed constraints can be
  imposed not only on atomic relations, but also on views. These features make
  our logic particularly suited for expressing inter-schema relationships in
  the context of information integration \cite{CDLNR98}, where it is crucial to
  be able to state that a certain concept of a schema corresponds (by means of
  inclusion or equivalence) to a view in another schema.
\item Finally, regular expressions can be profitably used to represent in the
  schema inductively defined structures such as sequences and lists, imposing
  complex conditions on them.
\end{itemize}

One of the distinguishing features of \DLRreg is that it is equipped with a
method for checking logical implication.  Indeed, $\DLRreg$ shares
EXPTIME-completeness of schema satisfiability and logical implication with many
expressive Description Logics \cite{CDLN01,BCMNP03} (see below).


We point out that \DLRreg supports only special forms of functional and
inclusion dependencies.  Hence the undecidability result of implication for
(general) functional and inclusion dependencies taken together, shown in
\cite{Mitc83,ChVa85}, does not apply.

\subsection{Queries}

A query $q$ for a \DLRreg schema is a non-recursive Datalog query, written in
the form:
\[
  q(\vett{x}) ~\la~ \conj_1(\vett{x},\vett{y}_1,\vett{c}_1) \lor\cdots\lor
                    \conj_m(\vett{x},\vett{y}_m,\vett{c}_m)
\]
where each $\conj_i(\vett{x},\vett{y}_i,\vett{c}_i)$ is a conjunction of
\emph{atoms}, and $\vett{x},\vett{y}_i$ (resp.~$\vett{c}_i$) are all the
variables (resp.~constants) appearing in the conjunction. Each atom has one of
the forms $C(t)$ or $\rR(\vett{t})$, where
\begin{itemize}
\item $t$ and $\vett{t}$ are constants or variables in
  $\vett{x},\vett{y}_i,\vett{c}_i$
\item $C$ and $\rR$ are respectively concepts and relations expressions over
  $\S$.
\end{itemize}
The number of variables of $\vett{x}$ is called the \emph{arity} of $q$, i.e.,
the arity of the relation denoted by the query $q$.

We observe that the atoms in the queries are arbitrary \DLRreg concepts and
relations, freely used in the assertions of the schema. This distinguishes our
approach with respect to \cite{DLNS98,LeRo96d}, where no constraints can be
expressed in the schema on the relations that appear in the queries.

Given an interpretation $\I$ of a schema $\S$, a query $q$ for $\S$ of arity
$n$ is interpreted as the set $\Int{q}$ of $n$-tuples $(\dd{o}{n})$, with each
$o_i\in\dom$, such that, when substituting each $o_i$ for $x_i$, the formula
\[
  \exists\vett{y}_1\per\conj_1(\vett{x},\vett{y}_1,\vett{c}_1)\lor\cdots\lor
  \exists\vett{y}_m\per\conj_m(\vett{x},\vett{y}_m,\vett{c}_m)
\]
evaluates to true in $\I$.

If $q$ and $q'$ are two queries (of the same arity) for $\S$, we say that $q$
is \emph{contained in} $q'$ wrt $\S$, denoted $\S\models q\incl q'$, if
$\Int{q} \incl \Int{q'}$ for every model $\I$ of $\S$. Given a \DLRreg schema
$\S$ and two queries for $\S$
\begin{align*}
  q(\vett{x})  &\la \conj_1(\vett{x},\vett{y}_1,\vett{c}_1) \lor\cdots\lor
                    \conj_m(\vett{x},\vett{y}_m,\vett{c}_m)\\
  q'(\vett{x}) &\la \conj'_1(\vett{x},\vett{y}'_1,\vett{c}'_1) \lor\cdots\lor
                    \conj'_{m'}(\vett{x},\vett{y}'_{m'},\vett{c}'_{m'})
\end{align*}
we have that $\S\models q\incl q'$ iff there is no model $\I$ of $\S$ such
that, when substituting suitable objects in $\dom$ for
$\vett{x},\vett{y}_1,\ldots\vett{y}_m$, the formula
\begin{align*}
  & (\conj_1(\vett{x},\vett{y}_1,\vett{c}_1) \lor\cdots\lor
    \conj_m(\vett{x},\vett{y}_m,\vett{c}_m)) \land{}\\
  & \lnot\exists\vett{z}_1\per\conj'_1(\vett{x},\vett{z}_1,\vett{c}'_1)
    \land\cdots\land
    \lnot\exists\vett{z}_{m'}\per
    \conj'_{m'}(\vett{x},\vett{z}_{m'},\vett{c}'_{m'})
\end{align*}
evaluates to true in $\I$.  In other words, $\S\models q\incl q'$ if and only
if there is no model of $\S$ that makes the formula
\begin{align*}
  & (\conjl[1] \lor\cdots\lor \conjl[m]) \land\\
  & \lnot\conjr[1] \land\cdots\land \lnot\conjr[m']
\end{align*}
true, where $\vett{a}$, $\dd{\vett{b}}{m}$ are Skolem constants, i.e.,
constants not appearing elsewhere for which the unique name assumption does not
hold.

\emph{Query containment} is the problem of checking whether $\S\models q\incl
q'$, where $\S$, $q$, and $q'$ are given as input.  \emph{Query satisfiability}
is the problem of checking whether a given query is interpreted as a non-empty
set in at least one model of a given schema.


\subsection{Example}

\begin{figure*}[tbp]
  \centering
  \input{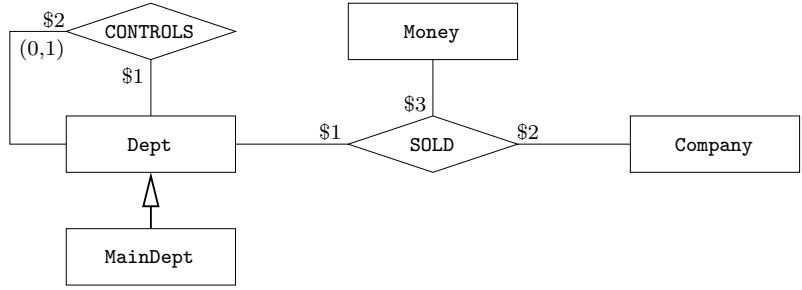}
  \caption{The entity-relationship diagram for the example in Section~2.3}
  \label{fig-ER-schema}
\end{figure*}

Consider an application where the departments of a given company can be
controlled by other departments, and sold to companies. Every department is
controlled by at most one department, and by at least one main department,
possibly indirectly. A main department is not controlled by any department.
If a main department is sold, then all the departments controlled by it are
also sold. Finally, if a department is sold, then all the department that,
directly or indirectly, controls it are also sold.

The basic concepts and relations are shown in Figure~\ref{fig-ER-schema} in the
form of an entity-relationship diagram. The specification of the application in
\DLRreg makes use of the concepts $\Dept$, $\MainDept$, $\Money$, $\Company$,
and the relations $\CONTROLS$, $\SOLD$. In particular, $\CONTROLS(x,y)$ means
that department $x$ has control over department $y$, and $\SOLD(x,y,z)$ means
that department $x$ has been sold to company~$y$ at price~$z$. The schema $\S$
is constituted by the following assertions:
\begin{align*}
  \SOLD     &\ISA \COMP{\$1}{\Dept} \AND \COMP{\$2}{\Company} \AND
                  \COMP{\$3}{\Money}\\
  \CONTROLS &\ISA \COMP{\$1}{\Dept} \AND \COMP{\$2}{\Dept}\\
  \Dept     &\ISA \ATMOSTR{1}{\$2}{\CONTROLS} \AND
                  \SOME{(\PROJ{\CONTROLS}{\$2}{\$1})^*}{\MainDept}\\
  \MainDept &\ISA \Dept \AND \NOT\SOMER{\$2}{\CONTROLS}\\
  \MainDept \AND \SOMER{\$1}{\SOLD} &\ISA
    \ALL{(\PROJ{\CONTROLS}{\$1}{\$2})^*}{\SOMER{\$1}{\SOLD}}\\
  \Dept \AND \SOMER{\$1}{\SOLD} &\ISA
    \SOME{(\PROJ{\CONTROLS}{\$2}{\$1})^*}{(\MainDept\AND\SOMER{\$1}{\SOLD})}
\end{align*}
The first two assertions are used to specify the types of the attributes of the
relations.
The third and the fourth assertions specify the basic properties of $\Dept$ and
$\MainDept$. It is easy to see that such assertions imply that, in all the
models of $\S$, the set of $\CONTROLS$ links starting from an instance $m$ of
$\MainDept$ form a tree (which we call $\CONTROLS$-tree) with root $m$.  The
role of the transitive closure $(\PROJ{\CONTROLS}{\$2}{\$1})^*$ and the number
restrictions is crucial for correctly representing the above property in the
schema.
Finally, the last two assertions, each one stating inclusions between views,
specify the company policy for selling departments.  Note again the use of the
transitive closure for this purpose.

We now consider two queries for the schema $\S$. The first query, called $q$ is
used to retrieve all the pairs of departments that are controlled by the same
department and that comprise at least one sold department. The second query,
called $q'$, retrieves all the pairs $(x,y)$ of departments such that $x$ has
been sold, and $y$ belongs to the same $\CONTROLS$-tree of $x$.  The queries
$q$ and $q'$ are defined as follows:
\begin{align*}
  q(x)  &\la \CONTROLS(x,y) \land \SOLD(y,z_1,z_2)\\
  q'(x) &\la \Dept(x) \land \SOLD(x,z_1,z_2)
\end{align*}
One can verify that $\S\models q\incl q'$.  Indeed, the schema $\S$ imposes
that (i)~the $\CONTROLS$ relation is typed, so that $x$ in $q$ is a department;
(ii)~when a department is sold, there is a main department (possibly
indirectly) controlling it that is also sold, and when a main department is
sold, all the departments it (directly and indirectly) controls are sold as
well.

Also, if we add to $q(x)$ the condition that department $x$ is not sold, we
obtain the query
\[
  q''(x) \la \CONTROLS(x,y) \land \SOLD(y,z_1,z_2) \land \NOT\SOLD(x,w_1,w_2)
\]
which is unsatisfiable.



\section{Checking Query Containment}
\label{sec-containment}

We address the problem of deciding, given a schema $\S$ and two queries $q$ and
$q'$ of the same arity, whether $\S\models q\incl q'$.  To do so, we make use
of a reduction of query containment to a problem of unsatisfiability in a
variant of Propositional Dynamic Logic, called \CPDLg.
In the next subsection, we introduce \CPDLg.  Then, we present the reduction,
prove its correctness, and analyze the computational complexity of the
resulting containment algorithm.

\subsection{The Propositional Dynamic Logic \CPDLg}
\label{sec-cpdlg}

Propositional Dynamic Logics are specific modal logics originally proposed as a
formal system for reasoning about computer program schemas \cite{FiLa79}. Since
then PDLs have been studied extensively and extended in several ways (see
e.g., \cite{KoTi90} for a survey).

Here, we make use of \CPDLg (studied in \cite{DeLe96} in the context of
description logics), which is an extension of Converse PDL \cite{KoTi90} with
\emph{graded modalities} \cite{FaDe85}. The syntax of \CPDLg is as follows
($A$ denotes an \emph{atomic formula}, $\phi$ an arbitrary \emph{formula}, $p$
an \emph{atomic program}, and $r$ an arbitrary \emph{program}):
\begin{align*}
  \phi &::= A ~\mid~
            \lnot\phi ~\mid~
            \phi_1\land\phi_2 ~\mid~
            \DIAM{r}{\phi} ~\mid~
            \QBOX{k}{p}{\phi} ~\mid~
            \QBOX{k}{p^-}{\phi}\\
  r    &::= p ~\mid~ r_1; r_2 ~\mid~ r_1\cup r_2 ~\mid~ r^* ~\mid~ \phi?
            ~\mid~ r^-
\end{align*}
We use the standard abbreviations, namely $\true$ for true, $\false$ for false,
$\lor$ for disjunction, $\limp$ for material implication, and $\BOX{r}{\phi}$
for $\lnot\DIAM{r}{\lnot\phi}$.

\begin{figure}[t]
  \boxtext{
   \[
     \begin{array}{rcl}
       \Int[\M]{A} &\subseteq& S\\
       \INT[\M]{\NOT \phi} &=& S\setminus\Int[\M]{\phi}\\
       \INT[\M]{\phi_1 \land \phi_2} &=& \Int[\M]{\phi_1}\cap\Int[\M]{\phi_2}\\
       \INT[\M]{\DIAM{r}{\phi}} &=& \set{s \mid\exists s'.
         (s,s') \in \Int[\M]{r} \land s' \in \Int[\M]{\phi}}\\
       \INT[\M]{\QBOX{k}{p}{\phi}} &=& \set{s \mid \card{\set{s' \mid
             (s,s') \in \Int[\M]{p} \land s'\in\Int[\M]{\phi}}} \leq k}\\
       \INT[\M]{\QBOX{k}{p^-}{\phi}} &=& \set{s \mid \card{\set{s' \mid 
             (s',s) \in \Int[\M]{p} \land s'\in\Int[\M]{\phi}}} \leq k}\\[3mm]
       \Int[\M]{p} &\subseteq& S\times S\\
       \INT[\M]{r_1; r_2}&=&\Int[\M]{r_1}\circ\Int[\M]{r_2}\\
       \INT[\M]{r_1\cup r_2}&=&\Int[\M]{r_1}\cup\Int[\M]{r_2}\\
       \INT[\M]{r^*}&=&(\Int[\M]{r})^* ~~=~~ \bigcup_{i\ge 0}(\Int[\M]{r})^i\\
       \INT[\M]{\phi?}&=& \set{(s,s) \mid s\in\Int[\M]{\phi}}\\
       \INT[\M]{r^-}&=& \set{(s,s') \mid (s',s)\in \Int[\M]{r}}
     \end{array}
   \]
  }
  \caption{Semantic rules for \CPDLg}
  \label{fig-pdlg-semantics}
\end{figure}

As usual for PDLs, the semantics of \CPDLg is based on \emph{Kripke
 structures} $\M=(S,\Int[\M]{\cdot})$, where $S$ is a \emph{set of states} and
$\Int[\M]{\cdot}$ is a mapping interpreting formulae as subsets of $S$ and
programs as binary relations over $S$. The semantics of each construct is
shown in Figure~\ref{fig-pdlg-semantics}.

It can be shown that \CPDLg has typical properties of PDLs, in particular the
\emph{connected-model property} (if a formula has a model, then it has one that
is connected when viewing it as a graph), the \emph{tree-model property} (if a
formula has a model, then it has one that is a tree when viewing it as an
undirected graph), and \emph{EXPTIME-completeness} of checking satisfiability
of a formula (with the assumption that numbers in graded modalities are
represented in unary) \cite{DeLe96,CDLN01,BCMNP03}.

\subsection{Reduction of Query Containment to Unsatisfiability in \CPDLg}
\label{sec-reduction}

Our aim is to reduce query containment to a problem of unsatisfiability in
\CPDLg.  To this end, we construct a \CPDLg formula starting from an instance
of the query containment problem. More precisely, if we have to check whether
there is no model of $\S$ that makes the formula
\[
  \begin{array}{l}
    (\conjl[1] \lor\cdots\lor \conjl[m]) \land{}\\
    \lnot\conjr[1] \land\cdots\land \lnot\conjr[m']
  \end{array}
\]
true, where $\vett{a}$, $\dd{\vett{b}}{m}$ are Skolem constants, we check the
unsatisfiability of the \CPDLg formula
\[
  \Phisqq = \Phi_{\S} \land
         (\bigvee_{j=1}^m\Phi_{\conj_j}) \land
         (\bigwedge_{j=1}^{m'}\lnot\Phi_{\conj'_j}) \land
         \Phiaux,
\]
constructed as described below.

\subsection*{$\Phi_\S$: encoding of $\S$}

\begin{figure}[t]
  \boxtext{
   \begin{minipage}{5cm}
     \[
       \begin{array}{rcl}
         \sigma(\top_1) &=& \top_1\\
         \sigma(A) &=& A\\
         \sigma(\NOT C) &=& \top_1\land\lnot\sigma(C)\\
         \sigma(C_1\AND C_2) &=& \sigma(C_1)\land\sigma(C_2)\\
         \sigma(\SOME{E}{C}) &=& \DIAM{\sigma(E)}{\sigma(C)}\\
         \sigma(\SOMER{\$i}{\rR}) &=& \DIAM{f_i^-}{\sigma(\rR)}\\
         \sigma(\ATMOSTR{k}{\$i}{\rR}) &=& \QBOX{k}{f_i^-}{\sigma(\rR)}
       \end{array}
     \]
   \end{minipage}
   \begin{minipage}{8cm}
     \[
       \begin{array}{rcl}
         \sigma(\top_n) &=& \top_n\\
         \sigma(\pP) &=& \pP\\
         \sigma(\COMP{i/n}{C}) &=& \top_n \land \BOX{f_i}{\sigma(C)}\\
         \sigma(\NOT\rR) &=& \top_n \land \lnot\sigma(\rR)\\
         \sigma(\rR_1\AND \rR_2) &=& \sigma(\rR_1) \land \sigma(\rR_2)\\[3mm]
         \sigma(\PROJ{R}{\$i}{\$j}) &=& f_i^-;\sigma(R)?;f_j\\
         \sigma(E_1\circ E_2) &=& \sigma(E_1);\sigma(E_2)\\
         \sigma(E_1\OR E_2) &=& \sigma(E_1)\cup\sigma(E_2)\\
         \sigma(E^*) &=& \sigma(E)^*
       \end{array}
     \]
   \end{minipage}

   \vspace{\abovedisplayskip}
   }
  \caption{Mapping $\sigma(\cdot)$ from \DLRreg to \CPDLg}
  \label{fig-mapping}
\end{figure}

$\Phi_{\S}$ is the translation of $\S$ into a \CPDLg formula, that is based
on \emph{reification} of $n$-ary relations, i.e., a tuple in a model of $\S$ is
represented in a model of $\Phisqq$ by a state having one functional link $f_i$
for each tuple component $\$i$.
$\Phi_{\S}$ makes use of the mapping $\sigma(\cdot)$ from \DLRreg expressions
to \CPDLg formulae defined in Figure~\ref{fig-mapping}.
The atomic formula $\top_1$ denotes those states that represent objects, while
each atomic formula $\top_n$, with $n\geq 2$, denotes those states that
represent tuples of arity $n$.
We denote with $\uu$ the program $(\create \cup f_1\cup\cdots\cup f_{n_{max}}
\cup \create^- \cup f_1^-\cup\cdots\cup f_{n_{max}}^-)^*$, where
$\create,f_1,\ldots,f_{n_{max}}$ are all atomic programs used in $\Phisqq$.
Due to the connected-model property of \CPDLg, $\uu$ represents the universal
accessibility relation.  Therefore, for a given interpretation,
$\BOX{\uu}{\phi}$ expresses that $\phi$ holds in every state, and
$\DIAM{\uu}{\phi}$ expresses that $\phi$ holds in some state.

$\Phi_{\S}$ is the conjunction of the following formulae:
\begin{align}
  & \BOX{\uu}{(\top_1\lor\cdots\lor\top_{n_{max}})}
    \label{for-reif-1}\\
  & \BOX{\uu}{\QBOX{1}{f_i}{\true}}
     & \text{for each $i\in\setone{n_{max}}$}
    \label{for-reif-2}\\
  & \BOX{\uu}{(\top_n \:\equiv\:
    \DIAM{f_1}{\top_1} \land\cdots\land \DIAM{f_n}{\top_1} \land
    \BOX{f_{n+1}}{\false})} \hspace*{-1.5cm}
    & \text{for each $n\in\setone[2]{n_{max}}$}
    \label{for-reif-3}\\
  & \BOX{\uu}{(\BOX{f_i}{\false} \:\limp\: \BOX{f_{i+1}}{\false})}
     & \text{for each $i\in\setone{n_{max}}$}
    \label{for-reif-4}\\
  & \BOX{\uu}{(A \:\limp\: \top_1)}
    & \text{for each atomic concept $A$}
    \label{for-reif-5}\\
  & \BOX{\uu}{(\pP \:\limp\: \top_n)}
    & \text{for each atomic relation $\pP$ of arity $n$}
    \label{for-reif-6}\\
  & \BOX{\uu}{(\sigma(C_1) \limp \sigma(C_2))}
    & \text{for each assertion $C_1\ISA C_2$ in $S$}
    \label{for-reif-7}\\
  & \BOX{\uu}{(\sigma(\rR_1)\limp\sigma(\rR_2))}
    & \text{for each assertion $\rR_1\ISA\rR_2$ in $S$}
    \label{for-reif-8}
\end{align}
The formula~(\ref{for-reif-1}) above expresses that each state represents
an object or a tuple of arity between $2$ and $n_{max}$.
The formula~(\ref{for-reif-2}) expresses that all programs $f_i$ are
functional (i.e., deterministic).
The formulae~(\ref{for-reif-3}) and~(\ref{for-reif-4}) express that the states
representing tuples of arity~$n$ are exactly those connected through programs
$f_1,\ldots,f_n$ to states representing objects, and not connected via programs
$f_i$, with $i>n$, to any state.
The formulae~(\ref{for-reif-5}) and~(\ref{for-reif-6}) express that states
satisfying atomic propositions corresponding to atomic concepts (resp.~atomic
relations of arity~$n$) are states representing objects (resp.~tuples of
arity~$n$).
Finally, the  formulae~(\ref{for-reif-7}) and~(\ref{for-reif-8}) encode the
assertions in $S$.


\subsection*{$\Phi_{\conj_j}$: encoding of each $\conjl$}

For each $j\in\setone{m}$, the encoding $\Phi_{\conj_j}$ of $\conjl$ makes use
of special atomic propositions, called \emph{name-formulae} whose
distinguishing properties are specified by $\Phiaux$ (see later).
Specifically, one name-formula $N_t$ is introduced for each term $t$ in
$\vett{a}$, $\vett{b}_j$, $\vett{c}_j$, and one name-formula $N_{\vett{t}}$ for
each tuple $\vett{t}$ such that for some $\rR$, $\rR(\vett{t})$ appears in
$\conjl$.
A name-formula assigns a name to a term~$t$ (resp.~tuple~$\vett{t}$), which
allows for identifying in a model certain states which correspond to $t$
(resp.~reified counterpart of~$\vett{t}$).  The distinguishing properties of
name-formulae guarantee that these states share some crucial properties that
allow us to isolate a single state as a representative of $t$
(resp.~$\vett{t}$).

Once we have name-formulae in place, we define $\Phi_{\conj_j}$ as the
conjunction of the following formulae:
\begin{enumerate}
\item \label{for-conj-1} for each name-formula $N_{\vett{t}}$ corresponding to
  a tuple $\vett{t}=(\dd{t}{n})$ appearing in $\conjl$
  \[
    \begin{array}{l}
      \BOX{\uu}{(N_{\vett{t}} \:\equiv\:
       \DIAM{f_1}{N_{t_1}}\land\cdots\land\DIAM{f_n}{N_{t_n}} \land
       \BOX{f_{n+1}}{\false})}\\[1mm]
      \BOX{\uu}{(N_{t_i} \:\limp\:
       (\DIAM{f_i^-}{N_{\vett{t}}} \land \QBOX{1}{f_i^-}{N_{\vett{t}}}))}
       \qquad \text{for each $i\in\{1,\ldots,n\}$}
    \end{array}
  \]
\item \label{for-conj-2} for each atom $C(t)$ in $\conjl$
  \[\BOX{\uu}{(N_{t} \:\limp\: \sigma(C))}\]
\item \label{for-conj-3} for each atom $\rR(\vett{t})$ in
  $\conj_j(\vett{a},\vett{b}_j,\vett{c}_j)$
  \[\BOX{\uu}{(N_{\vett{t}} \:\limp\: \sigma(\rR))}\]
\end{enumerate}
Intuitively, $\Phi_{\conj_j}$ expresses the relationships between terms and
tuples in $\conjl$ by using reification and name-formulae.  In particular, the
formulae~(\ref{for-conj-1}) relate the name-formulae corresponding to tuples
to the name-formulae corresponding to their components.  Each
formula~(\ref{for-conj-1}) and~(\ref{for-conj-2}) expresses that the states
satisfying the name-formula corresponding to a term (resp.~tuple) appearing in
an atom, satisfy also the formula corresponding to the predicate of the atom.

\subsection*{$\Phi_{\conj'_j}$: encoding of each $\conjr$}

Now consider a $j\in\setone{m'}$.  We construct the formula $\Phi_{\conj'_j}$
as a disjunction of formulae, one for each possible partition of the variables
$\vett{z}_j$ in $\conjr$.  More precisely, to build one such formula, we
consider a partition $\pi$ of the variables $\vett{z}_j$.  Then, for each
equivalence class in the partition we choose a variable as a representative,
and substitute in $\conjr$ all other variables in the same equivalence class by
the representative, thus obtaining a formula $\conjpart$.  Now, from such a
formula we build a corresponding \CPDLg formula by making use of a special
graph, called \emph{tuple-graph}, which intuitively reflects the dependencies
between variables and tuples resulting from the appearance of the variables in
the atoms of $\conjpart$%
\footnote{The tuple-graph is similar to the graph used in \cite{ChRa97} to
 detect cyclic dependencies between variables.}. %
A tuple-graph is a directed graph with nodes labeled by \CPDLg formulae and
edges labeled by \CPDLg programs, formed as follows:
\begin{itemize}
\item There is one node $t$ for each term $t$ in $\vett{a}$, $\vett{w}$,
  $\vett{c}_j$, and one node $\vett{t}$ for each tuple $\vett{t}$ such that
  $\rR(\vett{t})$ appears in $\conjpart$.
  Each node $t$ is labeled by all $\sigma(C)$ such that $C(t)$ appears in
  $\conjpart$.
  Each node $\vett{t}$ is labeled by all $\sigma(\rR)$ such that
  $\rR(\vett{t})$ appears in $\conjpart$.
\item There is one edge labeled by $f_i$ from the node $\vett{t}=(\dd{t}{n})$
  to the node $t_i$, $i\in\setone{n}$, for each tuple $\vett{t}$ such that
  $\rR(\vett{t})$ appears in $\conjpart$. %
\end{itemize}

Notice that dividing the variables $\vett{z}_j$ in $\conjr$ in all possible
ways into equivalence classes and replacing equivalent variables by one
representative, corresponds to introducing in all possible ways equalities
between variables.  Such equalities allow us to take into account that a cycle
in the tuple graph can in fact be eliminated, and become simply a chain, when
different variables are assigned the same object.  As will become clear in the
following, the distinction between variables appearing in cycles in the
tuple-graph and those that do not, is indeed necessary for the correctness of
the proposed technique for query containment under constraints.


In the following, we call \emph{formula-template} a \CPDLg formula in which
formula-placeholders occur that later will be substituted by actual formulae.
From the tuple-graph of $\conjpart$ we build a \CPDLg formula-template
$\deltat$, and to do so we have to consider that in general the tuple-graph is
composed of several connected components.  For the $i$-th connected component
we build a formula-template $\deltat[i]$ by choosing a starting node $t_0$
(corresponding to a term) and performing a depth-first visit of the
corresponding component and building the formula in a postorder fashion.  We
describe the construction by defining a visiting function $V$, which, given a
node of the tuple-graph, returns the corresponding formula-template, and as a
side effect marks the nodes of the graph that it visits.
\begin{itemize}
\item If $u=t$, then $V(t)$ marks $t$, and returns the conjunction of:
  \begin{enumerate}
  \item[(i)] $t$ itself, used as a placeholder, and every formula labeling the
    node $t$;
  \item[(ii)] for each edge $(\vett{t},t)$ labeled by $f_i$ (i.e., $t=t_i$ in
    $\vett{t}$) such that $\vett{t}$ is not marked yet, the formula
    $\DIAM{f_i^-}{V(\vett{t})}$.
  \end{enumerate}
\item If $u=\vett{t}=(\dd{t}{n})$, then $V(\vett{t})$ marks $\vett{t}$, and
  returns the conjunction of:
  \begin{enumerate}
  \item[(i)] $\vett{t}$ itself, used as a placeholder, and every formula
    labeling the node $\vett{t}$;
  \item[(ii)] for each edge $(\vett{t},t_i)$ labeled by $f_i$, such that $t_i$
    is not marked yet, the formula $\DIAM{f_i}{V(t_i)}$;
  \item[(iii)] for each edge $(\vett{t},t_i)$ labeled by $f_i$, such that $t_i$
    is already marked, the formula $\DIAM{f_i}{t_i}$.
  \end{enumerate}
\end{itemize}
Then the formula-template $\deltat[i]$ for the $i$-th connected component is
defined as $V(t_0)$, where $t_0$ is the starting node chosen for the visit.

The formula-template $\deltat$ for the whole tuple-graph of $\conjpart$,
composed of $\ell\geq 1$ connected components, is
\[
  \DIAM{\uu}{\deltat[1]} \land\cdots\land \DIAM{\uu}{\deltat[\ell]}
\]
where $\deltat[1],\ldots,\deltat[\ell]$ are the formula-templates corresponding
to all the connected components in the tuple-graph of $\conjpart$.

Now we are ready to define the \CPDLg formula $\varphi_{\pi}$ corresponding
to a partition $\pi$ of the variables $\vett{z}_j$.  The formula
$\varphi_{\pi}$ consists of the disjunction of all formulae obtained by
replacing in the formula-template $\deltat$
\begin{enumerate}
\item[(i)] each placeholder $\vett{t}$ by $\top_n$ where $n$ is the arity of
  the tuple $\vett{t}$;
\item[(ii)] each placeholder $d$ in $\vett{a}$, $\vett{c}_j$ by the
  name-formula $N_d$;
\item[(iii)] each placeholder $w_i$ corresponding to a variable not occurring
  in a cycle in the tuple-graph by $\top_1$;
\item[(iv)] each placeholder $w_i$ corresponding to a variable occurring in a
  cycle in the tuple-graph by each of the name-formulae $N_t$ corresponding to
  a term in $\vett{a}$, $\dd{\vett{b}}{m}$, $\dd{\vett{c}}{m}$ occurring in $q$
  or to a term in $\dd{\vett{c}'}{m'}$ occurring in $q'$.
\end{enumerate}
Observe that the number of such disjuncts in $\varphi_\pi$ is
$O(\ell_1^{\ell'_2})$, where $\ell_1$ is the number of variables and constants
in $q$ plus the number of constants in $q'$, and $\ell'_2$ is the number of
variables $w_i$ occurring in a cycle in the tuple-graph for $\conjpart$.

Since $\varphi_{\pi}$ corresponds to one possible partition of the variables
$\vett{z}_j$, we obtain the formula $\Phi_{\conj'_j}$ as the disjunction of all
formulae $\varphi_{\pi}$, one for each possible partition $\pi$ of the
variables $\vett{z}_j$.  The number of such disjuncts is $O(2^{\ell_2})$, where
$\ell_2$ is the number of variables $\vett{z}_j$.

Therefore, the total number of disjuncts for $\Phi_{\conj'_j}$ is
$O(\ell_1^{O(\ell_2)})$.

\subsection*{$\Phiaux$: encoding of constants and variables}

Let $\Phi' = \Phi_{\S} \land (\bigvee_{j=1}^m\Phi_{\conj_j}) \land
(\bigwedge_{j=1}^{m'}\lnot\Phi_{\conj'_j})$, and let $\dd{N}{K}$ be all
name-formulae in $\Phi'$.  $\Phiaux$ is formed by the conjunction of:
\begin{itemize}
\item the formula %
  $\DIAM{\create}{N_1}\land\cdots\land\DIAM{\create}{N_K}$ which expresses
  the existence of a state satisfying a name-formula $N_i$, for each
  $i\in\setone{K}$;
\item one formula of the form %
  $\BOX{\uu}{(N_{c_i} \limp \lnot N_{c_j})}$ for each pair of distinct
  constants $c_i$, $c_j$ appearing in the queries (not Skolem constants);
\item one formula of the form %
  $\BOX{\uu}{(N_i\land \phi \:\limp\: \BOX{\uu}{(N_i\limp\phi)})}$ for
  each name-formula $N_{i}$, $i\in\{1,\ldots,K\}$, %
  and each formula $\phi$ such that%
  \footnote{$\CL(\phi)$ is the Fisher-Ladner closure of a \CPDLg formula
   $\phi$, and $\Pre(r)$ is the set of ``prefixes'' of a program $r$
   \cite{DeLe96}.}:
  \begin{enumerate}
  \item[(a)] $\phi\in\CL(\Phi')$,
  \item[(b)] $\phi=\DIAM{\ol{r}}{\phi'}$ with $\DIAM{r}{\phi'}\in\CL(\Phi')$,
    and
  \item[(c)] $\phi=\DIAM{\ol{r}';p}{N_j}$ with $r'\in\Pre(r)$, ~$p=f \mid f^-$,
    and $r$, $f$, $N_j$ occurring in $\CL(\Phi')$
  \end{enumerate}
  where $\ol{r}$ is defined inductively as follows:
  \begin{align*}
    \ol{p}           &= p;(\land_i \lnot N_i)?\\
    \ol{r_1;r_2}     &= \ol{r_1};\ol{r_2}\\
    \ol{r_1\cup r_2} &= \ol{r_1}\cup\ol{r_2}\\
    \ol{r_1^*}       &= \ol{r_1}^*\\
    \ol{\phi?}       &= \phi?
  \end{align*}

\end{itemize}

The role of $\Phiaux$ is to enforce that, in every model of $\Phisqq$, for each
$N_k$, one representative state can be singled out among those satisfying
$N_k$.  This would be trivially obtained if we could force all these states to
satisfy exactly the same formulae of the logic. $\Phiaux$ forces a weaker
condition, namely that these states satisfy the same formulae in the finite set
(whose size is polynomial with respect to $\Phi'$) described above.
Theorem~\ref{thm-unique-object-model} shows that this is sufficient for our
purposes.

\begin{figure*}[tb]
  \centering
  \begin{minipage}[c]{10cm}
    \input{model.pstex_t}
  \end{minipage}
  \hspace{1cm}
  \begin{minipage}[c]{4cm}
    \begin{tabular}{lcl}
      $\Phisqq$ &is true in& $s_{root}$\\
      $N_{\vett{t}}\land p$ &is true in& $s_{\vett{t}}$\\
      $N_{a_1}$ &is true in& $s_{a_1}$\\
      $N_{a_2}$ &is true in& $s_{a_2}$
    \end{tabular}
  \end{minipage}
  \caption{A model of $\Phisqq$}
  \label{fig-model}
\end{figure*}

\bigskip

We illustrate the encoding of the containment problem $\S\models q\incl q'$
into unsatisfiability of the \CPDLg formula $\Phisqq$ by means of the following
example.

\begin{example} \label{exa-encoding}
  Consider two queries
  \begin{align*}
    q(x_1,x_2)  &\la p(x_1,x_2)\\
    q'(x_1,x_2) &\la r(x_1,x_2,z)
  \end{align*}
  over a schema $\S$ such that $\S\not\models q\incl q'$.
  Figure~\ref{fig-model} schematically shows a model of the formula $\Phisqq$
  that represents a counterexample to the containment.  Indeed, the model
  contains a state $s_{\vett{t}}$ in which $p$ holds, that, being connected to
  $s_{a_1}$ and $s_{a_2}$ by means of $f_1$ and $f_2$, respectively, represents
  a tuple $(a_1,a_2)$ that satisfies $p$. Since $s_{a_1}$ and $s_{a_2}$ satisfy
  $N_{a_1}$ and $N_{a_2}$, respectively, and
  $\Phi_{\conj'}=\BOX{U}{(N_{a_1}\limp\BOX{f_1^-}{(r\limp\BOX{f_2}{(\lnot
     N_{a_2}\lor\BOX{f_3}{\false})})})}$ is true in $s_{root}$, it follows that
  $s_{a_1}$ satisfies $\BOX{f_1^-}{(r\limp\BOX{f_2}{\lnot N_{a_2}})}$.
  Therefore, in the model there is no state satisfying $r$ representing a tuple
  $(a_1,a_2,z)$.
\end{example}

\subsection{Correctness of the Reduction}
\label{sec-correctness}

By exploiting the properties of the encoding $\Phisqq$, we can now prove
decidability of query containment in our case.

We say that a tuple-graph $g$ is \emph{satisfied} in an interpretation
$\M$ for $\Phisqq$ if there exists an homomorphism $\eta$ mapping the nodes of
$g$ to states of $\M$ such that:
\begin{itemize}
\item if a node $u$ of $g$ is a (possibly Skolem) constant, then
  $\eta(u)\in\Int[\M]{N_{u}}$;
\item if a node $u$ of $g$ is labeled by a formula $\phi$, then
  $\eta(u)\in\Int[\M]{\phi}$;
\item if an edge $(u,u')$ of $g$ is labeled by a program $f$ then
  $(\eta(u),\eta(u'))\in\Int[\M]{f}$.
\end{itemize}

Given a formula-template $\phi$ and a substitution $\theta$ of its
placeholders, we denote by $\phi\theta$ the formula obtained from $\phi$ by
substituting the placeholders according to $\theta$.

\begin{lemma}\label{thm-tuple-graph}
  Let $g$ be a connected component of a tuple-graph, $\delta$ the corresponding
  formula-template, and $\M$ a \CPDLg interpretation.  If there exists a
  substitution $\theta$ of the placeholders such that $\INT[\M]{\delta\theta}$
  is not empty, then $g$ is satisfied in $\M$.
\end{lemma}
\begin{proof}
If $\INT[\M]{\delta\theta}$ is not empty then it is possible to define an
homomorphism as follows.  Let $s_t$ (resp.~$s_{\vett{t}}$) be the state of $\M$
that is used in satisfying $\delta\theta$ in the position corresponding to $t$
(resp.~$\vett{t}$), then $\eta(t)=s_t$ (resp.~$\eta(\vett{t})=s_{\vett{t}}$).
\end{proof}

\begin{theorem} \label{thm-soundness}
  Let $\S$ be a schema, $q$, $q'$ two queries of the same arity, and $\Phisqq$
  the formula obtained as specified above.  If $\S\not\models q \subseteq q'$
  then $\Phisqq$ is satisfiable.
\end{theorem}

\begin{proof}
It suffices to consider a model $\I=(\dom,\Int{\cdot})$ of $\S$ that makes the
following formula true:
\begin{align*}
  & (\conjl[1] \lor\cdots\lor \conjl[m]) \land{}\\
  & \lnot\conjr[1] \land\cdots\land \lnot\conjr[m']
\end{align*}

From $\I$ build a reified \CPDLg interpretation $\M=(S,\Int[\M]{\cdot})$ for
$\Phisqq$ as follows:
\begin{itemize}
\item $S = \dom \cup \set{s_{root}} \cup
  \bigcup_{n\in\set{2,\ldots,\nmax}}
  \set{s_{\vett{t}}\mid \vett{t}\in\Int{\top_n}}$;
\item for each $n\in\set{2,\ldots,\nmax}$, for each $(t_1,\ldots,t_n) \in
 \Int{\top_n}$, we have $s_{(t_1,\ldots,t_n)} \in\Int[\M]{\top_n}$, and
 $(s_{(t_1,\ldots,t_n)},t_i)\in\Int[\M]{f_i}$ with $i\in\setone{n}$;
\item for each atomic relation $\pP$, for each $(t_1,\ldots,t_n) \in
  \Int{\pP}$, we have $s_{(t_1,\ldots,t_n)} \in\Int[\M]{\pP}$
\item $\Int[\M]{\top_1} = \dom$ and for each atomic concept $A$, we have
  $\Int[\M]{A}=\Int{A}$;
\item for each (possibly Skolem) constant $t$ occurring in $\conjl[1]
  \lor\cdots\lor \conjl[m]$ we have $\Int[\M]{N_t} = \set{t}$; similarly for
  each tuple $\vett{t}$ of (Skolem) constants occurring in $\conjl[1]
  \lor\cdots\lor \conjl[m]$ we have $\Int[\M]{N_{\vett{t}}} =
  \set{s_{\vett{t}}}$;
\item for each (possibly Skolem) constant $t$ occurring in $\conjl[1]
  \lor\cdots\lor \conjl[m]$ we have $(s_{root},t)\in\Int[\M]{\create}$;
  similarly for each tuple $\vett{t}$ of (Skolem) constants occurring in
  $\conjl[1] \lor\cdots\lor \conjl[m]$ we have
  $(s_{root},s_{\vett{t}})\in\Int[\M]{\create}$.
\end{itemize}

Next we show that $\M$ is a model of the formula $\Phisqq$.
It is immediate to verify that
\begin{enumerate}
\item $s_{root}\in\Int[\M]{\Phi_\S}$ (by construction, considering that $\I$ is
  a model of $\S$);
\item $s_{root}\in\Int[\M]{\Phi_{\conj_j}}$, for some $j\in\setone{m}$ (by
  construction, considering that $\I$ satisfies
  $\conjl[1]\lor\cdots\lor\conjl[m]$);
\item $s_{root}\in\Int[\M]{\Phiaux}$ (by construction, considering that
  name-formulae are interpreted as singletons in $\M$).
\end{enumerate}
It remains to show that $s_{root}\not\in\Int[\M]{\Phi_{\conj'_j}}$, for each
$j\in\setone{m'}$.  Suppose not, that is, suppose that
$s_{root}\in\Int[\M]{\Phi_{\conj'_j}}$, for some $j\in\setone{m'}$.  Then
there exists a partition $\pi$ of the variables $\vett{z}_j$ in $\conjr[j]$
such that $s_{root}\in\Int{\varphi_\pi}$. This in turn implies that there is a
substitution $\theta$ of the placeholders in the formula-template
$\DIAM{\uu}{\deltat[1]}\land\cdots\land\DIAM{\uu}{\deltat[\ell]}$ such that
$s_{root}\in\Int[\M]{(\DIAM{\uu}{\deltat[1]}\land\cdots\land
 \DIAM{\uu}{\deltat[\ell]})\theta}$.  But then we have
$s_{root}\in\INT[\M]{(\DIAM{\uu}{\deltat[1]})\theta \land\cdots\land
 (\DIAM{\uu}{\deltat[\ell]})\theta}$, i.e., for each $i\in\setone{\ell}$, there
is a state $s_i\in\INT[\M]{\deltat[i]\theta}$.
By Lemma~\ref{thm-tuple-graph} this implies that, for each $i\in\setone{\ell}$,
the connected component  $g_i$ corresponding to $\deltat[i]$ of the tuple-graph
is satisfied in $\M$, and hence, by construction, the corresponding part of
$\conjr[j]$ is satisfied in $\I$. Since this is true for all connected
components, we get that the whole $\conjr[j]$ is satisfied in $\I$,
contradicting the fact that $\I$ makes $\lnot\conjr[j]$ true.
\end{proof}

We say that a model of $\Phisqq$ is \emph{tuple-admissible} if there is no pair
of states that represent the same reified tuple.
We say that a model of $\Phisqq$ is \emph{admissible} if it is tuple-admissible
and each name-formula is true in exactly one state.
We say that a model $\M=(S,\Int[\M]{\cdot})$ of $\Phisqq$ is a
\emph{pseudo-tree admissible model} if it is admissible and has the following
form:
\begin{itemize}
\item it has a distinguished state $s_{root}$, and $K$ not necessarily distinct
  states $s_{N_1},\ldots,s_{N_K}$, one for each name-formula $N_i$, such that
  $\Int[\M]{N_i}=\set{s_{N_i}}$;
\item $\Int[\M]{\create}=\set{(s_{root},s_{N_i}) \mid i\in\setone{K}}$;
\item each maximal connected component of $\M \setminus (\set{s_{root}} \cup
  \set{s_{N_i} \mid i\in\setone{K}})$ is a tree, when viewed as an undirected
  graph.
\end{itemize}
Notice that, the subgraph induced by $\M\cap\set{s_{N_i}\mid i\in\setone{K}}$
is an arbitrary graph, instead.

The following theorem shows that, w.r.t.\ satisfiability, one can restrict the
attention to pseudo-tree admissible models.

\begin{theorem} \label{thm-unique-object-model}
  Let $\S$ be a schema, $q$, $q'$ two queries of the same arity, and $\Phisqq$
  the formula obtained as specified above.  If $\Phisqq$ is satisfiable then it
  has a pseudo-tree admissible model.
\end{theorem}

\begin{proof}
By the tree-model property, $\Phisqq$ admits a tree-model
$\M=(S,\Int[\M]{\cdot})$, in which obviously there is no pair of states that
represent the same reified tuple.  Let $s_{root}\in\Int[\M]{\Phisqq}$ be the
root of $\M$.
We transform $\M$ into a new model $\M'=(S',\Int[\M']{\cdot})$ with $S'\incl
S$, which interprets name-formulae as singletons and is still
tuple-admissible, as follows.  For each $N_i$, $i\in\setone{K}$, we select a
state $s_{N_i}$, among the states $s\in\Int[\M]{N_i}$ such that
$(s_{root},s)\in\Int[\M]{\create}$.  Then we define:
\begin{align*}
  \Int[\M']{\create} &= \set{(s_{root},s_{N_i})\in\Int[\M]{\create} \mid
                        i\in\setone{K}}\\
  \Int[\M']{p} &=
    \begin{array}[t]{@{}l}
      (\Int[\M]{p} \setminus (
        \begin{array}[t]{@{}l}
          \set{(s_{N_i},s)\in\Int[\M]{p} \mid
             s\in\Int[\M]{N_j}, i,j\in\setone{K}} \cup{}\\
          \set{(s,s_{N_j})\in\Int[\M]{p} \mid
             s\in\Int[\M]{N_i}, i,j\in\setone{K}}))
         \end{array}\\
        {}\cup \set{(s_{N_i},s_{N_j}) \mid
          (s_{N_i},s)\in\Int[\M]{p}, s\in\Int[\M]{N_j}, i,j\in\setone{K}}\\
        \qquad\text{for each atomic program $p$ except $\create$}
     \end{array}\\
    \Int[\M']{N_i} &= \set{s_{N_i}}
      \qquad\text{for each name-formula $N_i$, $i\in\setone{K}$}\\
    \Int[\M']{A} &= \Int[\M]{A}\cap S'
      \qquad\text{for each atomic formula $A$ except name formulae}\\
    S' &= 
          \set{s_{root}} \cup
          \set{s\in S \mid (s_{root},s)\in
               \Int[\M']{\create} \circ
               (\bigcup_p(\Int[\M']{p}\cup{\INT[\M']{p^-}}))^*}
\end{align*}
It is possible to show, by using the construction in Lemma~5 of \cite{DeLe96}%
\footnote{The construction in \cite{DeLe96} is phrased in the Description Logic
 \CIQ, and it is used to reduce ABox reasoning to satisfiability. \CIQ and
 \CPDLg can be seen as a syntactic variant one of the other, and our handling
 of constants, through name-formulae, in \CPDLg is closely related to handling
 ABoxes in \CIQ, the only difference is that for constants in the ABoxes the
 unique name assumption is made, while here we do not make such an assumption.
 However, the unique name assumption plays no role in the construction of
 \cite{DeLe96}, hence that construction works in our case as well.},%
that for each $\phi\in\CL(\Phisqq)$ and for each state $s\in\S'$
\[s\in\Int[\M']{\phi} \text{ if and only if } s\in\Int[\M]{\phi}.\]
Hence, since $\Phisqq\in\CL(\Phisqq)$ and $s_{root}\in\INT[\M]{\Phisqq}$, we
get the thesis.
\end{proof}

For pseudo-tree admissible models one can prove the ``converse'' of
Lemma~\ref{thm-tuple-graph}.
\begin{lemma}\label{thm-tuple-graph-pseudo-tree}
  Let $g$ be a tuple-graph, $\phi$ the corresponding formula-template, and $\M$
  a \CPDLg pseudo-tree admissible model of $\Phisqq$.  Then we have: if there
  exists an homomorphism $\eta$ from $g$ to $\M$ such that nodes corresponding
  to variables in $g$ are mapped either to states representing (possibly
  Skolem) constants or to distinct states, then there exists a substitution
  $\theta$ of the placeholders in $\phi$ such that $\INT[\M]{\phi\theta}$ is
  not empty.
\end{lemma}
\begin{proof}
We first observe that, since $\M$ is a pseudo-tree admissible model, and $\eta$
assigns all variables in $g$ not assigned to states representing (Skolem)
constants, to distinct states, we have that, if a variable $w$ occurs in a
cycle in $g$, then the state $\eta(w)$ assigned to $w$ must be one representing
a (possibly Skolem) constant.

Hence we can define $\theta$ as the substitution that:
\begin{itemize}
\item replaces each placeholder that corresponds to a variable $w$ occurring on
  a cycle in $g$, and thus such that $\eta(w)$ is a (possibly Skolem) constant,
  with a name formula $N_d$.
\item replaces each placeholder that corresponds to a variable $w$ not
  occurring on a cycle in $g$, with $\top_1$.
\end{itemize}
It is easy to verify that, with $\theta$ defined in this way
$\INT[\M]{\phi\theta}$ is not empty.
\end{proof}

\begin{theorem} \label{thm-completeness}
  Let $\S$ be a schema, $q$, $q'$ two queries of the same arity, and $\Phisqq$
  the formula obtained as specified above.  If $\Phisqq$ has a pseudo-tree
  admissible model then $\S\not\models q\incl q'$.
\end{theorem}

\begin{proof}
We show how to construct from a pseudo-tree admissible model $\M$ of $\Phisqq$
a model $\I$ of $\S$ in which there is a tuple $\vett{a}$ of objects such that
$\vett{a}\in\Int{q}$ and $\vett{a}\not\in\Int{q'}$.
$\I$ is built as follows:
\begin{itemize}
\item $\dom = \Int[\M]{\top_1}$;
\item $\Int{\pP} = \set{(\dd{s}{n}) \mid \exists s'\in\Int[\M]{\pP}\per
   ((s',s_i)\in\Int[\M]{f_i}, \text{ for } i\in\setone{n})}$,
  for each atomic relation $\pP$ of arity $n$;
\item $\Int{A} = \Int[\M]{A}$, for each atomic concept $A$;
\item $\Int{t} = s\in\Int[\M]{N_t}$, for each constant and Skolem constant $t$
  in $q$ and $q'$.
\end{itemize}
To show that $\I$ does the job, we have to show that:
\begin{enumerate}
\item $\I$ is a model of $\S$;
\item $\conjl[1]\lor\cdots\lor\conjl[m]$ is true in $\I$, i.e., there is one
  $j\in\setone{m}$ such that $\conjl[j]$ is true in $\I$;
\item $\conjr[1]\land\cdots\land\conjr[m']$ is true in $\I$, i.e., for each
  $j\in\setone{m'}$, we have that $\conjr[j]$ is true in $\I$.
\end{enumerate}

To show that $\I$ is a model of $\S$ we can exploit the fact that
$\M=(S,\Int{\cdot})$ is a model of $\Phi_{\S}$ and that, since it is
admissible, there is no pair of states in $S$ that represent the same reified
tuple.  By construction of $\I$ it is easy to see that all assertions in $\S$
are true in $\I$.

To show that there is one $j\in\setone{m}$ such that $\conjl[j]$ is true in
$\I$, we exploit that $\M$ is an admissible model of $\Phisqq$.  Hence there is
a $j\in\setone{m}$ such that $\M$ is an admissible model of $\Phi_{\conj_j}$,
and since each name-formula is true in exactly one state, the claim easily
follows.

It remains to show that for each $j\in\setone{m'}$, we have that $\conjr[j]$ is
true in $\I$.  We show that, if for some substitution $\vett{o}=(\dd{o}{n})$
for the variables $\vett{z}_j=(\dd{z}{n})$ we have that
$\conj'_j(\vett{a},\vett{o},\vett{c}_j)$ is true in $\I$, then, for some
$j'\in\setone{m'}$, we get a contradiction to $\M$ is a model of
$\lnot\Phi_{\conj'_j}$.

By considering which variables have been assigned to the same objects in
$\vett{o}$, we get a partition of the variables in $\vett{z}_j$.  Corresponding
to such a partition $\pi$ we have considered in the construction of
$\Phi_{\conj'_j}$ the formula $\conjpart$, obtained by replacing all variables
in the same equivalence class by a representative.  Observe that, as a result,
distinct variables in $\vett{w}_\pi$ are assigned distinct objects in
$\vett{o}$.

Let now $\varphi_\pi$ be the disjunct in $\Phi_{\conj'_j}$ obtained from
$\conjpart$.  $\varphi_\pi$ is a disjunction of formulae, all obtained by
replacing in the same formula-template
$\DIAM{\uu}{\deltat[1]}\land\cdots\land\DIAM{\uu}{\deltat[\ell]}$ the
placeholders corresponding to the variables $\vett{w}$ either by $\top_1$ or by
name-formulae corresponding to constants or Skolem constants.

Let $g_i$ be the tuple-graph obtained from $\conjpart$.  Then, using the
assignment above we can define an homomorphism $\eta$, mapping the nodes of
$g_i$ to states of $\M$, such that nodes corresponding to variables in $g_i$
are either mapped to states representing (possibly Skolem) constants or mapped
to distinct states.
Hence, we can apply Lemma~\ref{thm-tuple-graph-pseudo-tree}, and conclude that
there exists a substitution $\theta$ of the corresponding formula-template
$\DIAM{\uu}{\deltat[1]}\land\cdots\land\DIAM{\uu}{\deltat[\ell]}$ such that
$(\DIAM{\uu}{\deltat[1]})\theta\land\cdots\land(\DIAM{\uu}{\deltat[\ell]})\theta$
is true in $\M$. This implies that one of the disjuncts in $\varphi_\pi$ is
true in $\M$ and hence that $\lnot\Phi_{\conj'_j}$ is false in $\M$. Thus we
get a contradiction.
\end{proof}

The following theorem, which is a consequence of Theorems~\ref{thm-soundness},
\ref{thm-unique-object-model} and~\ref{thm-completeness}, shows decidability of
query containment under constraints in our setting.

\begin{theorem} \label{thm-soundness-completeness}
  Let $\S$ be a schema, $q$, $q'$ two queries of the same arity, and $\Phisqq$
  the formula obtained as specified above.  Then $\S\not\models q\incl q'$ if
  and only if $\Phisqq$ is satisfiable.
\end{theorem}

\subsection{Complexity of Query Containment}
\label{sec-complexity}

We analyze now the computational complexity of our algorithm for query
containment.

\begin{theorem} \label{thm-upper-bound-general}
  Let $\S$ be a schema and $q$ and $q'$ two queries.  Then deciding whether
  $\S\models q\subseteq q'$ can be done in time
  $2^{p(|\S|+|q|+|q'|\cdot\ell_1^{\ell_2})}$, where $|\S|$, $|q|$, and $|q'|$
  are respectively the sizes of $\S$, $q$, and $q'$, $\ell_1$ is the sum of the
  number of variables in $q$ and the number of constants in $q$ and $q'$, and
  $\ell_2$ is the number of existentially quantified variables in $q'$.
\end{theorem}

\begin{proof}
Soundness and completeness of the encoding of query containment $\S\models
q\subseteq q'$ into unsatisfiability of $\Phisqq$ follow from
Theorem~\ref{thm-soundness-completeness}.  With regard to complexity, since
satisfiability in \CPDLg is EXPTIME-complete, it follows that query containment
can be done in time $2^{p(|\Phisqq|)}$.  It is easy to verify that
$|\Phisqq|=O(|\S|+|q|+|q'|\cdot\ell_1^{O(\ell_2)})$.
\end{proof}

The previous theorem provides, for query containment $\S\models q\subseteq q'$,
a single exponential upper bound in the size of $\S$ and of $q$, and a double
exponential upper bound in the size of $q'$ (note that $|q'|$ is an upper bound
for $\ell_2$).
The single exponential upper bound in the size of $\S$ and of $q$ is tight.
Indeed, it follows from EXPTIME-hardness of satisfiability in \CPDLg (in fact
plain PDL \cite{FiLa79}) and from the fact that any \CPDLg formula can be
expressed as a \DLRreg concept.  EXPTIME-hardness in $\S$ holds even in the
case where $\S$ does not contain regular expressions.  Indeed, the formulae
used in the EXPTIME-hardness proof of satisfiability in PDL \cite{FiLa79}, can
be expressed as assertions in \DLRreg not involving regular expressions.
It is still open whether the double-exponential upper bound in the size of $q'$
is tight.

The double exponential upper bound in the size of $q'$ is due to the
exponential blowup in the size of $\Phisqq$.  By analyzing the reduction
presented in Section~\ref{sec-reduction}, one can observe that such an
exponential blowup is only due to those existentially quantified variables in
$q'$ that appear inside a cycle in the tuple-graph for $q'$.  Hence, when the
tuple-graph for $q'$ does not contain cycles, we have that
$|\Phisqq|=O(|\S|+|q|+|q'|)$, and query containment can be checked in time
$2^{p(|\S|+|q|+|q'|)}$.
A relevant case when this occurs is when (the tuple-graph for) the query on the
right-hand side has the structure of a tree.

\begin{corollary} \label{thm-upper-bound-tree}
  Let $\S$ be a schema, $q$ and $q'$ two queries of the same arity, and let
  $q'$ have the structure of a tree.  Then deciding whether $\S\models
  q\subseteq q'$ can be done in time $2^{p(|\S|+|q|+|q'|)}$.
\end{corollary}

Observe that this gives us an EXPTIME-completeness result for containment of an
arbitrary query in a tree-structured one wrt a schema.

Query satisfiability can be considered as a special case of query containment.
Indeed, given a schema $\S$, a query $q$ is satisfiable wrt $\S$ if and only if
it is not contained in the empty query wrt $\S$.  The empty query can be
expressed, for example, as
$u(\vett{x})\la\pP(\vett{x})\land\lnot\pP(\vett{x})$, where $\vett{x}$ is a
tuple of variables and $\pP$ is a new atomic relation, both of the same arity
as $q$.

\begin{corollary} \label{thm-upper-bound-satisfiability}
  Let $\S$ be a schema, and $q$ a query.  Then deciding whether $q$ is
  satisfiable wrt $\S$ can be done in time $2^{p(|\S|+|q|)}$.
\end{corollary}

Again, this result shows EXPTIME-completeness of query satisfiability wrt a
schema.



\section{Undecidability of Containment of Queries with Inequalities}
\label{sec-undecidability}

In this section we show that, if we allow for inequalities inside the queries,
then query containment wrt a schema becomes undecidable.
The proof of undecidability exploits a reduction from the unbounded
\emph{tiling problem} \cite{Emde97}.  An instance $\T=(\D,H,V)$ of the tiling
problem is defined by a finite set $\D$ of tile types, a horizontal adjacency
relation $H\in\D\times\D$, and a vertical adjacency relation $V\in\D\times\D$,
and consists in determining whether there exists a tiling of the first quadrant
of the integer plane with tiles of type in $\D$ such that the adjacency
conditions are satisfied.
As shown in \cite{Hare85,Emde97}, the tiling problem is well suited to show
undecidability of variants of modal and dynamic logics, and the difficult part
of the proof usually consists in enforcing that the tiles lie on an integer
grid.  To this end we exploit a query containing one inequality.

Formally, given an instance $\T=(\D,H,V)$ of the tiling problem, a
\emph{$\T$-tiling} is a total function $t:\Nat\times\Nat\lora\D$, and such a
tiling is \emph{correct} if $(t(i,j),t(i+1,j))\in H$ and $(t(i,j),t(i,j+1))\in
V$, for each $i,j\in\Nat$.  We reduce the problem of checking whether there
exists a correct $\T$-tiling to the problem of checking whether $\S_{\T}\models
q_0\subseteq q'_0$, for suitable schema $\S_{\T}$ and queries $q_0$ and $q'_0$
containing inequalities.

Consider an instance $\T=(\D,H,V)$ of the tiling problem with tile types
$\D=\{\dd{D}{k}\}$.  We construct a schema $\S_\T$ using the atomic concepts
$\Tile$, $\dd{D}{k}$ and two binary atomic relations $\Right$ and $\Up$ as
follows:
\begin{align}
  \Tile &~\ISA~ D_1 \OR\cdots\OR D_k
    \label{eqn-tiling-union-1}\\
  D_i   &~\ISA~ \Tile \qquad\mbox{for each $i\in\setone{k}$}
    \label{eqn-tiling-union-2}\\
  D_i   &~\ISA~ \NOT D_j \qquad\mbox{for each $i,j\in\setone{k}$, $i<j$}
    \label{eqn-tiling-disjoint}\\
  \Tile &~\ISA~ \ATMOSTR{1}{\$1}{\Right} \AND \ATMOSTR{1}{\$1}{\Up}
    \label{eqn-tiling-functional}\\
  \Tile &~\ISA~ \SOMER{\$1}{(\Right\AND\COMP{\$2}{\Tile})} \AND
                \SOMER{\$1}{(\Up\AND\COMP{\$2}{\Tile})}
    \label{eqn-tiling-existence}\\
  D_i   &~\ISA~ \textstyle{
               \begin{array}[t]{@{}l}
                 (\bigsqcup_{(D_i,D_j)\in H}
                    \NOT\SOMER{\$1}{(\Right\AND\COMP{\$2}{\NOT D_j})}) ~\AND\\
                 (\bigsqcup_{(D_i,D_j)\in V}
                    \NOT\SOMER{\$1}{(\Up\AND\COMP{\$2}{\NOT D_j})})
                 \qquad\mbox{for each $i\in\setone{k}$}
               \end{array}}
    \label{eqn-tiling-adjacency}
\end{align}
The define the boolean queries $q_0$ and $q'_0$ as follows:
\begin{align*}
  q_0()  &\la \Tile(x)\\
  q'_0() &\la \Right(x,y)\land \Up(y,z)\land
              \Up(x,y')\land \Right(y',z')\land
              z\neq z'
\end{align*}

\begin{theorem} \label{thm-tiling-reduction}
  Let $\T$ be an instance of the tiling problem, $\S_{\T}$ a schema, and $q_0$
  and $q'_0$ two queries defined as specified above.  Then there is a correct
  $\T$-tiling if and only if $\S_{\T}\not\models q_0\incl q'_0$.
\end{theorem}

\begin{proof}
\OnlyIf %
Let $t$ be a correct $\T$-tiling.  We construct an interpretation $\I_t$ of
$\S_\T$ as follows:
\begin{align*}
  \dom[\I_t] &= \Nat\times\Nat\\
  \Int[\I_t]{\Tile} &=\dom[\I_t]\\
  \Int[\I_t]{\D_h} &=\{(i,j)\in\dom[\I_t] \mid t(i,j)=D_h\},\qquad \mbox{for
   each $h\in\{1,\ldots,k\}$}\\
  \Int[\I_t]{\Right} &=\{((i,j),(i+1,j)) \mid i,j\in\Nat\}\\
  \Int[\I_t]{\Up}    & =\{((i,j),(i,j+1)) \mid i,j\in\Nat\}
\end{align*}
It is immediate to verify that $\I_t$ is a model of $\S_{\T}$ and that
$\Int[\I_t]{q_0}$ is true while $\Int[\I_t]{{q'_0}}$ is false.

\If %
Consider a model $\I$ of $\S_\T$ in which $q_0$ is true and $q'_0$ is false.
Then $\I$ contains an instance $o_0$ of $\Tile$ and
assertions~(\ref{eqn-tiling-existence}) in $\S_\T$ force the existence of
arbitrary long chains of instances of $\Tile$, beginning with $o_0$ and
connected one to the next by alternations of $\Int{\Right}$ and $\Int{\Up}$.
By assertions~(\ref{eqn-tiling-functional}), $\Right$ and $\Up$ are functional
for all instances of $\Tile$, and since $q'_0$ is false in $\I$, these chains
of objects form indeed a grid.  By assertions~(\ref{eqn-tiling-union-1})
and~(\ref{eqn-tiling-disjoint}), each such object is an instance of precisely
one $D_h$.  Hence, we can construct a tiling $t_{\I}$ by assigning to each
object $o$ of the grid, representing an element of the first quadrant, a unique
tile type $D_h$.  Considering also assertions~(\ref{eqn-tiling-adjacency}), it
is easy to show by induction on the length of the chain from $o_0$ to an
instance $o$ of $\Tile$, that the horizontal and vertical adjacency conditions
for $o$ are satisfied.  Hence $t_{\I}$ is a correct $\T$-tiling.
\end{proof}

The theorem above immediately implies undecidability of containment wrt a
schema of queries containing inequalities.

\begin{theorem} \label{thm-undecidability}
  Let $\S$ be a schema, and $q$, $q'$ two queries of the same arity that may
  contain atoms of the form $t\neq t'$.  Then the query containment problem
  $\S\models q\incl q'$ is undecidable.
\end{theorem}

The reduction used in the proof of Theorem~\ref{thm-tiling-reduction} shows
that query containment remains undecidable even in the restricted case where:
\begin{itemize}
\item $\S$ does not contain assertions on relations, and all assertions on
  concepts are of the form $A\ISA C$,
\item $\S$, $q$, and $q'$ do not contain regular expressions,
\item $q$ and $q'$ do not contain union, or constants expressions, and
\item there is a single inequality in $q'$, and no inequality in $q$.
\end{itemize}

Making use of a more involved proof, it is possible to show that the reduction
used in Theorem~\ref{thm-tiling-reduction} works also if one omits from
$\S_{\T}$ assertions~(\ref{eqn-tiling-functional}) specifying functionality of
$\Right$ and $\Up$.  In this case, a model $\I$ of $\S_{\T}$ in which $q_0$ is
true and $q'_0$ is false does no longer determine a unique grid, but it is
nevertheless possible to extract from $\I$ a correct $\T$-tiling.



\section{Query Answering}
\label{sec-answering}

As we said in the introduction, it is well known in the database
literature that there is a tight connection between the problems of
conjunctive query containment and conjunctive query answering
\cite{ChMe77}.  Such a relationship has had a particular importance in
settings of databases with incomplete information, such as those
arising in information integration \cite{AbDu98,Lenz02},
semistructured data \cite{CDLV02b}, and Description Logics
\cite{BCMNP03}.
In this section we discuss query answering under Description Logics
constraints, taking advantage of the results on query containment presented
above.  By query answering under Description Logics constraints we mean to
compute the answers to a query over an \emph{incomplete database}, i.e., a
database that is partially specified and must satisfy all Description Logic
constraints expressed in a schema.%
\footnote{Note that, the case in which we have complete information on the
 database, the constraints do not play any role on query answering, assuming
 that the database is consistent with them.}

Given a $\DLRreg$ schema $\S$, we specify an incomplete database $\D$ over $\S$
by means of a set of facts, called \emph{membership assertions}, of the form
\[
  C(a) \qquad\qquad \rR(\vett{a})
\]
where $C$ and $\rR$ are respectively a concept expression and a relation
expression over $\S$, $a$ is a constant, and $\vett{a}$ is an tuple of
constants of the same arity as $\rR$.
Note that such a notion of incomplete database corresponds to that of ABox in
Description Logics \cite{BCMNP03}.

An interpretation $\I$ satisfies an assertion $C(a)$ if $\Int{a}\in\Int{C}$,
and it satisfies an assertion $\rR(\vett{a})$ if $\Int{\vett{a}}\in\Int{\rR}$.
We say that $\I$ is a \emph{model} of $\D$, if it satisfies all assertions in
$\D$.
An incomplete database $\D$ is \emph{satisfiable with respect to a schema} $\S$
if there is an interpretation $\I$ that is a model of both $\S$ and $\D$.
Intuitively, every such interpretation $\I$ represents a complete database that
is coherent with both $\D$, and the Description Logic constraints in $\S$.

Given a schema $\S$, an incomplete database $\D$ over $\S$, and a query $q$ for
$\S$, the set of \emph{certain answers} $\cert{q}{\S}{\D}$ of $q$ with respect
to $\S$ and $\D$ is the set of tuples $\vett{c}$ of constants in $\D$ that are
answers to $q$ for all complete databases coherent with $\D$ and $\S$, i.e.,
such that $\vett{c}\in\Int{q}$, for all models $\I$ of $\S$ and $\D$.

Given a query
\[
  q(\vett{x}) ~\la~ \conj_1(\vett{x},\vett{y}_1,\vett{c}_1) \lor\cdots\lor
                    \conj_m(\vett{x},\vett{y}_m,\vett{c}_m)
\]
in order to check whether a tuple $\vett{c}$ of constants is in
$\cert{q}{\S}{\D}$, we can resort to query containment \cite{AbDu98}.  In
particular, let us define the boolean (i.e., of arity 0) queries $Q_{\D}$ and
$Q_{q,\vett{c}}$ as follows:
\[
  \begin{array}{lcl}
    Q_{\D}() &\la& \bigwedge_{C(a)\in\D} C(a) \land
       \bigwedge_{\rR(\vett{a})\in\D} \rR(\vett{a})\\
    Q_{q,\vett{c}}() &\la& \conj_1(\vett{c},\vett{y}_1,\vett{c}_1)
       \lor\cdots\lor \conj_m(\vett{c},\vett{y}_m,\vett{c}_m)
  \end{array}
\]
The first query $Q_{\D}$ is the conjunction of all facts in $\D$, while the
second query $Q_{q,\vett{c}}$ is obtained from $q$ by replacing each variable
in $\vett{x}$ with the corresponding constant in $\vett{c}$.

\begin{theorem} \label{thm-qa-to-qc}
  Let $\S$ be a schema, $\D$ an incomplete database over $\S$, $q$ a query for
  $\S$, and $\vett{c}$ a tuple of constants in $\D$ of the same arity as $q$.
  Then $\vett{c}\in\cert{q}{\S}{\D}$ if and only if $\S\models Q_{\D}\subseteq
  Q_{q,\vett{c}}$.
\end{theorem}

\begin{proof}
The result can be proved exactly as in \cite{AbDu98}.
\end{proof}

From Theorem~\ref{thm-upper-bound-general} we immediately obtain the following
complexity result.

\begin{theorem} \label{thm-upper-bound-answering}
  Let $\S$ be a schema, $\D$ an incomplete database over $\S$, $q$ a query for
  $\S$, and $\vett{c}$ a tuple of constants in $\D$ of the same arity as $q$.
  Then deciding whether $\vett{c}\in\cert{q}{\S}{\D}$ can be done in time
  $2^{p(|\S|+|\D|+|q|\cdot d^{\ell})}$, where $|\S|$, $|\D|$, and $|q|$ are
  respectively the sizes of $\S$, $\D$, and $q$, $d$ is the number of constants
  in $\D$ and $q$, and $\ell$ is the number of existentially quantified
  variables in $q$.
\end{theorem}

Note that this means that, while query answering is double exponential in
combined complexity, it is actually only single exponential in the number of
constants in the database.  It follows, that our technique is exponential in
data complexity, i.e., the complexity measured only with respect to the size of
$\D$.

Finally, it follows directly from the semantics, that satisfiability of a given
incomplete database $\D$ with respect to a schema $\S$.  can be rephrased as
satisfiability of the query $Q_{\D}$ with respect to $\S$.  Thus, we obtain the
following result.

\begin{corollary} \label{thm-upper-bound-TBox-ABox}
  Let $\S$ be a schema and $\D$ an incomplete database over $\S$.  Then
  deciding whether $\D$ is satisfiable with respect to $\S$ can be done in time
  $2^{p(|\S|+|\D|)}$.
\end{corollary}

In Description Logics jargon, this shows EXPTIME-completeness of TBox+ABox
satisfiability in our setting.  Observe that, since we allow for union of
conjunctive queries on the left-hand side query in the containment, this result
can be immediately extended to satisfiability of a TBox together with a
\emph{disjunction} of ABoxes \cite{CaDL01}.



\section{Conclusions}
\label{sec-conclusions}

In this paper we have introduced \DLRreg, an expressive language for
specifying database schemas and non-recursive Datalog queries, and we
have presented decidability (with complexity) and undecidability
results of both the problem of checking query containment, and the
problem of answering queries under the constraints expressed in the
schema.

The query language considered in this paper allows no form of
recursion, not even the transitive closure of binary relations. It is
our aim in the future to extend our analysis to the case where queries
may contain regular expressions, in the spirit of \cite{CDLV00b}.



\bibliographystyle{acmtrans}
\bibliography{main-bib}

\begin{received}
Received July 2005
\end{received}

\end{document}